\documentclass{appolb}
\usepackage[latin2]{inputenc}
\usepackage[dvips]{graphicx}
\usepackage[mathcal]{eucal}
\usepackage{amsfonts}
\usepackage{amsmath}
\usepackage{amsthm}
\title{Periodic diffraction patterns for 1D quasicrystals} 

\author{ 
\textsc{Pawe\l{} Buczek} 
\footnote{Current address: Department of Physics, Avadh Bhatia Physics Laboratory, University of Alberta,
Edmonton T6G 2J1, Alberta, Canada (also address for correspondence).}
\footnote{Electronic address: \texttt{\small{pbuczek@phys.ualberta.ca}}}\\ 
\textit{\small{Faculty of Physics and Applied Computer Science, AGH-UST}} \\ 
\textit{\small{al. Mickiewicza 30, 30-059 Krak\'ow, Poland}} \\
\textsc{Lorenzo Sadun} 
\footnote{Electronic address: \texttt{\small{sadun@math.utexas.edu}}} \\
\textit{\small{Department of Mathematics, University of Texas}} \\
\textit{\small{1 University Station C1200, Austin, TX 78703, USA}} \\
\textsc{Janusz Wolny}
\footnote{Electronic address: \texttt{\small{wolny@novell.ftj.agh.edu.pl}}} \\
\textit{\small{Faculty of Physics and Applied Computer Science, AGH-UST}} \\ 
\textit{\small{al. Mickiewicza 30, 30-059 Krak\'ow, Poland}}
}

\date{} 

\begin{document}
\maketitle

\begin{sloppy}

\begin{abstract}
A simple model of 1D structure based on a Fibonacci sequence with variable atomic spacings is proposed. The model
allows for observation of the continuous transition between periodic and non-periodic diffraction patterns. The
diffraction patterns are calculated analytically both using ``cut and project" and ``average unit cell'' method,
taking advantage of the physical space properties of the structure.

\end{abstract}

\section{Introduction}
\label{sec:Introduction}
For nearly 100 years, the analysis of diffraction patterns of solids has been an essential tool for studying
solids, since the diffraction pattern of a solid is essentially the (squared) Fourier transform of the set of
atomic positions. Classical crystallography considered periodic structures, whose diffraction patterns consist
entirely of sharp Bragg peaks. The Fourier transform of such a periodic set can be computed from the relevant
unit cell. The discovery of quasicrystals showed that discrete diffraction patterns are associated not only with
periodic structures but also with a large family of solids that have no discrete translation symmetry --
quasicrystals. This fact was incorporated into a new definition of ``crystal" proposed in 1992 by the Commission
on Aperiodic Crystals established by the International Union of Crystallography: a crystal is defined to be any
solid with an essentially discrete diffraction pattern.

Diffraction patterns for periodic and aperiodic crystals differ in substantial ways -- for instance, the
diffraction patterns of quasicrystals may exhibit ``forbidden symmetry". It is therefore illuminating to consider
a model that interpolates between periodic and aperiodic structures, and observe how the diffraction pattern
changes.

We consider such a one-parameter family of structures in this paper.  Specifically, we consider a fixed
(Fibonacci) sequence of two types of ``atoms'', and vary the amount of space around each type of atom, while
keeping the overall density fixed. The control parameter $\kappa$ is the ratio of the two allowed distances
between nearest neighbors. In all cases, the diffraction pattern is discrete, and the locations of the Bragg
peaks are independent of $\kappa$.  However, the {\it intensities} of the peaks are $\kappa$-dependent.  When
$\kappa$ is rational, the intensities form a periodic pattern, while when $\kappa$ is irrational, the diffraction
pattern is aperiodic. We compute this diffraction pattern in two independent but equivalent ways: a) by
recovering periodicity going to higher dimension (the ``cut and project method" deeply discussed in many papers:
\cite{deBruijn1981, Duneau1985, Elser1986, Hof1995, Hof1997, Jagodzinski1991, Janssen1998, Jaric1986,
Kalugin1985, Kramer1984, Senechal1997}); b)using the concept of the reference lattice.

These results are in accordance with the ergodic theory of tiling spaces.  It is known that the Bragg peaks of a
tiling $T$ occur at eigenvalues of the generator of translations on the hull of $T$ (i.e., the space of all
tilings in the same local isomorphism class as $T$) \cite{Dworkin1993}. It is also known \cite{Radin2001} that
the hulls of modified Fibonacci chains with the same average spacing are topologically conjugate, hence that
their generators of translations have the same spectral decomposition. The question of when and how such a
modification affects the dynamical spectrum was addressed for one dimensional patterns in \cite{ClarkSadun2003a},
and for higher dimensional patterns in \cite{ClarkSadun2003b}. (It should be noted that for a substitution tiling
whose substitution matrix has two of more eigenvalues greater than 1, a generic change in tile length will
destroy the Bragg peaks altogether, in sharp contrast to the behavior of modified Fibonacci chains, other Pisot
substitutions, and other Sturmian sequences.)

Ergodic theory says nothing, however, about the intensities of the Bragg peaks.  Although the spectrum of the
generator of translations is complicated, for special values of the control parameter some of the peaks may have
intensity zero, resulting in a simpler diffraction pattern. The calculations in this paper demonstrate that this
does in fact happen.

\section{The Modified Fibonacci Chain}
\label{sec:RelevantPropertiesOfFibonacciSequence}
The properties of Fibonacci sequences have already been thoroughly studied (see e.g., \cite{Senechal1995}). They
are sequences of two elements $A$ and $B$ obtained from a substitution rule:
\begin{align}
A \longrightarrow AB; \quad B \longrightarrow A. 
\end{align}
Let $\vec{p}_{m}=(p_{m}^{A},p_{m}^{B})$ be the population vector, where $p_{m}^{X}$ tells how many elements of
type $X$ are among the first $m$ terms of the sequence. Of course, $p_{m}^{A}+p_{m}^{B}=m$. There are an
uncountably infinite number of Fibonacci sequences, but all have the same local properties and the same
diffraction pattern. It is easy to see that every Fibonacci sequence has
\begin{align}
  \lim_{m \to \infty}\frac{p_{m}^{A}}{p_{m}^{B}}=\tau, \quad
  \textrm{where} \quad  \tau=\frac{1+\sqrt{5}}{2}. \label{eq:conctr}
\end{align}
For definiteness, we will work with the sequence (\cite{Senechal1995})
\begin{align}
\vec{p}_{m}=(\left\|\frac{m}{\tau}\right\|,m-\left\|\frac{m}{\tau}\right\|).
\label{eq:3}
\end{align}
Here $\left\|\cdot\right\|$ is the nearest integer function: If $m \in \mathbb{Z}$ and $m \leq x < m+1$ then:
\begin{align}
\left\|x\right\|=
\left\{ 
\begin{array}{ll}
m & \textrm{if} \quad x \in [m,m+1/2), \\
m+1 & \textrm{if} \quad x \in [m+1/2,m+1).
\end{array}\right. \label{eq:nif}
\end{align}

Now pick two positive numbers (also called $A$ and $B$) that determine the space between each ``$A$'' or ``$B$''
atom and its predecessor.  That is, the atomic positions are given by
\begin{equation}
x_m = \vec p_m \cdot (A,B). \label{eq:positions}
\end{equation} 
We call the sequence $\{ x_m \}$ a {\it modified Fibonacci chain}. If $A=B$, then the atoms are equally spaced,
and this is simply a periodic array. If $A/B = \tau$, then the atomic positions are those obtained from the
canonical ``cut and project'' method.  By varying the ratio $A/B$, we interpolate between these two cases.  Note
that the average atomic spacing is $\frac{\tau A + B}{1+\tau}$. We will keep this average spacing fixed and
consider the one-parameter family
\begin{equation}
A = \tau - \frac{\epsilon}{\tau}, \qquad B = 1 + \epsilon,
\label{eq:a}
\end{equation}
depending on the control parameter $\epsilon$. 
 
\section{2D Analysis of the Modified Fibonacci Chain}
\label{sec:MFCh2D}
The modified Fibonacci chain can be obtained by a ``cut and project'' method with a nonstandard projection.  
From this construction we can compute the diffraction pattern.

Let $\mathcal{L}_{2}^{\nu}$ be the 2 dimensional square lattice with spacing $\nu$. The Vorono\"\i{} cell of each
lattice point is a square.  Let $l_{0}$ be the line $y=x/\tau$ through the origin, making an angle $\beta_0 =
\cot^{-1}(\tau)$ with the $x$-axis. Let $X$ be the subset of $\mathcal{L}_{2}^{\nu}$ whose Vorono\"\i{} cells are
cut by $l_{0}$. It is well known (see e.g., \cite{Senechal1995}) that $X = \left \{ \vec{x}\mid \vec{x}=\nu
  \vec{p}_{m}, m \in \mathbb{Z} \right\}$ where $\vec{p}_{m}$ are
population vectors given by (\ref{eq:3}).

\begin{figure}[f]
        \begin{center}
                \includegraphics[width=0.75\textwidth,origin=c, angle=0]{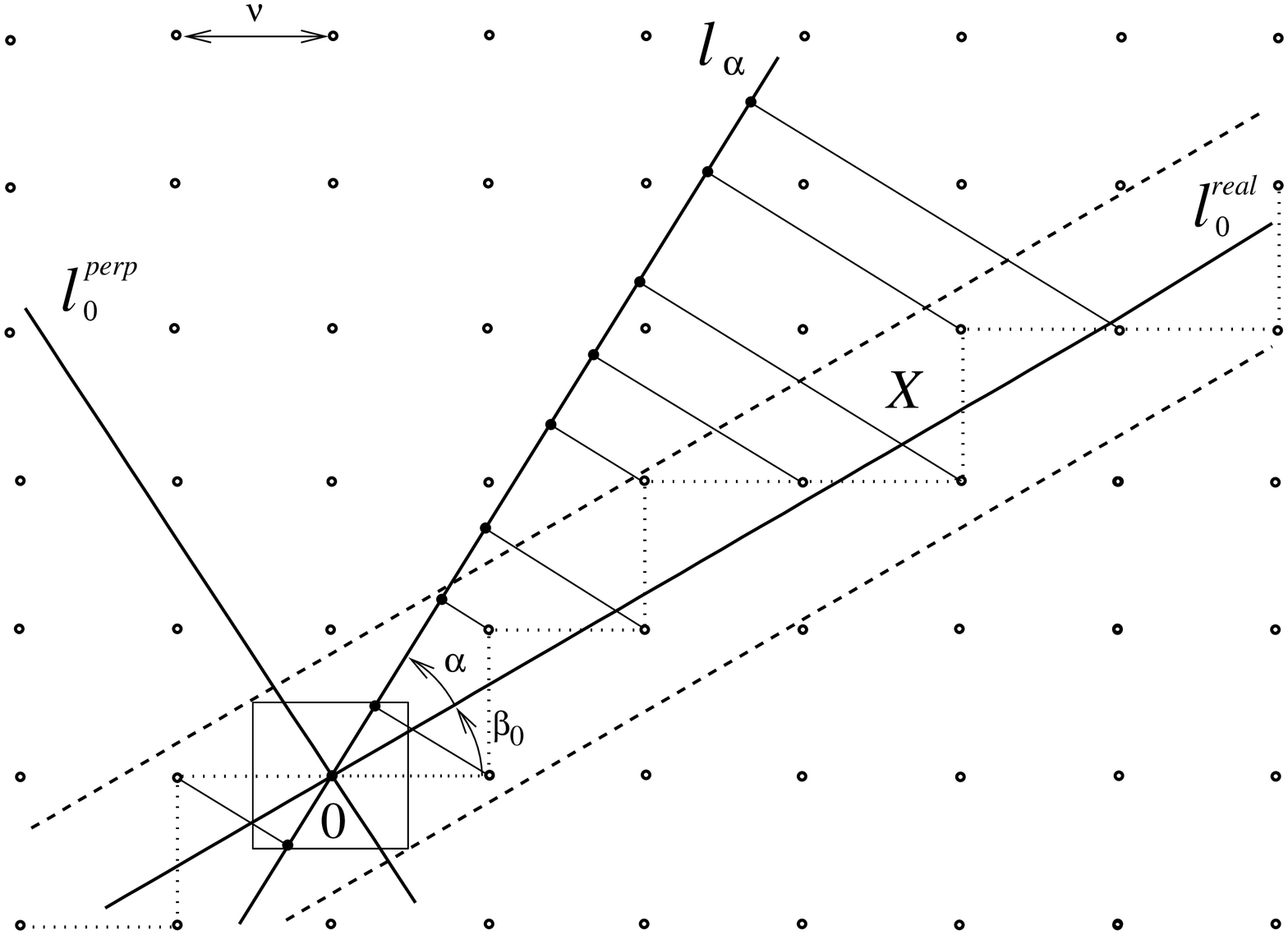}
        \end{center}
        \caption{The 2D construction of the modified Fibonacci chain. Details in text.}
        \label{fig:projection}
\end{figure}

Let $l_\alpha$ be the line through the origin making an angle
\begin{align}
\beta=\beta_{0}+\alpha  \label{eq:beta}
\end{align}
with the $x$-axis, and define $\Pi_{\alpha}$ to be the orthogonal projection onto $l_{\alpha}$. Finally, let
$\Lambda$ be the projection of $X$ onto $l_{\alpha}$:
\begin{equation}
\Lambda=\Pi_{\alpha}(X).
\label{eq:4}
\end{equation}
The set $\Lambda$ is then a modified Fibonacci chain; the procedure was presented in the figure \ref{fig:projection}. The two distances are
\begin{equation}
A=\nu \cos \beta, \qquad B=\nu \sin \beta. \label{eq:ABang}
\end{equation}
Their sequence is fully determined by $X$ and does not depend on $\alpha$. The average distance between nearest
neighbors in $\Lambda$ is
\begin{align}
\lim_{m \to \infty} \frac{\vec{p}_{m} \cdot (A,B)}{\vec{p}_{m} \cdot (1,1)} 
=\nu \frac{\sqrt{\tau+2}}{\tau+1} \cos \alpha.
\label{eq:6}
\end{align}
To keep this average spacing constant we take
\begin{align}
\nu = \frac{\sqrt{\tau +2}}{\cos \alpha}. \label{eq:nu}
\end{align}
The angle $\alpha$, the displacement parameter $\epsilon$ of (\ref{eq:a}) and the ratio $\kappa=A/B$ are related
by
\begin{align}
\epsilon = & \frac{\tau - \kappa}{\kappa + \tau -1} =
\tau \tan (\alpha), &\\
\kappa = \cot(\beta) = & \frac{\tau - \tan(\alpha)}{1 + \tau \tan(\alpha)} =
\frac{\tau + \epsilon (1-\tau)}{1 + \epsilon}, & \\
\tan(\alpha) = 
& \frac{\epsilon}{\tau} = \frac{\tau - \kappa}{\kappa \tau + 1}.&
\end{align}

We assume that every point of the set $\Lambda$ is an atom with scattering power equal to unity. Our aim is to
calculate the diffraction pattern of such a structure.  We begin by calculating the 2-dimensional diffraction
pattern of $X$.  The diffraction pattern of $\Lambda$ is then a section of the diffraction pattern of $X$ along
the direction $l_{\alpha}$ (figure \ref{fig:RecSpace}).

To get the diffraction pattern of $X$ we note that $X$ is $\mathcal{L}_{2}^{\nu}$ times the characteristic
function of a strip of width
\begin{align}
h_{\alpha} =\frac{\tau + 1}{\cos \alpha} 
\label{eq:ha}
\end{align}
around $l_0$.  The Fourier transform of a product is the convolution of the Fourier transforms, and the Fourier
transform of a lattice is the reciprocal lattice.  The diffraction pattern of $X$ in 2D has normalized intensity
\begin{equation}
I(\vec{k})=\sum_{m_{x}}\sum_{m_{y}}\Big(\frac{\sin((h_{\alpha}| \vec{k}-\vec{k}_{m_{x}m_{y}} |)/2)}{(h_{\alpha}|
\vec{k}-\vec{k}_{m_{x}m_{y}} |)/2)}\Big)^{2} \delta\big((\vec{k}-\vec{k}_{m_{x}m_{y}}) \cdot (\tau,1)\big),
\label{eq:9}
\end{equation}
where $\vec{k}_{m_{x}m_{y}}=\frac{2\pi}{\nu}(m_{x},m_{y})$ and $m_{x},\:m_{y}\in\mathbb{Z}$ label points of the
reciprocal lattice to $\mathcal{L}_{2}^{\nu}$.  Along the direction $l_\alpha$, the peaks can be observed at
positions
\begin{equation}
k^{phys}=\frac{\vec{k}_{m_{x}m_{y}} \cdot (\tau,1)}{\cos \alpha \sqrt{\tau +2}}
= 2 \pi \frac{\tau m_x + m_y}{\tau + 2}. \label{eq:10}
\end{equation}
Note that these positions are independent of $\alpha$ (or $\epsilon$ or $\kappa$). Their intensities are
\begin{equation}
I_{m_{x}m_{y}}=\Big(\frac{\sin w}{w}\Big)^{2}, \qquad w= \frac{h_\alpha \vec{k}_{m_{x}m_{y}} \cdot (-1-\tau \tan
\alpha, \tau - \tan \alpha)} {2 \sqrt{\tau + 2}}.
\label{eq:11}
\end{equation}
After simplifying and rewriting in terms of $\epsilon$ we obtain
\begin{equation}
w= \frac{\pi(\tau + 1)}{\tau + 2} 
(-m_{x}(1+\epsilon)+m_{y}(\tau - \epsilon / \tau)).  
\label{eq:intens_2D}
\end{equation}

\begin{figure}[f]
        \begin{center}
        \includegraphics[width=0.75\textwidth,
        angle=0.0] {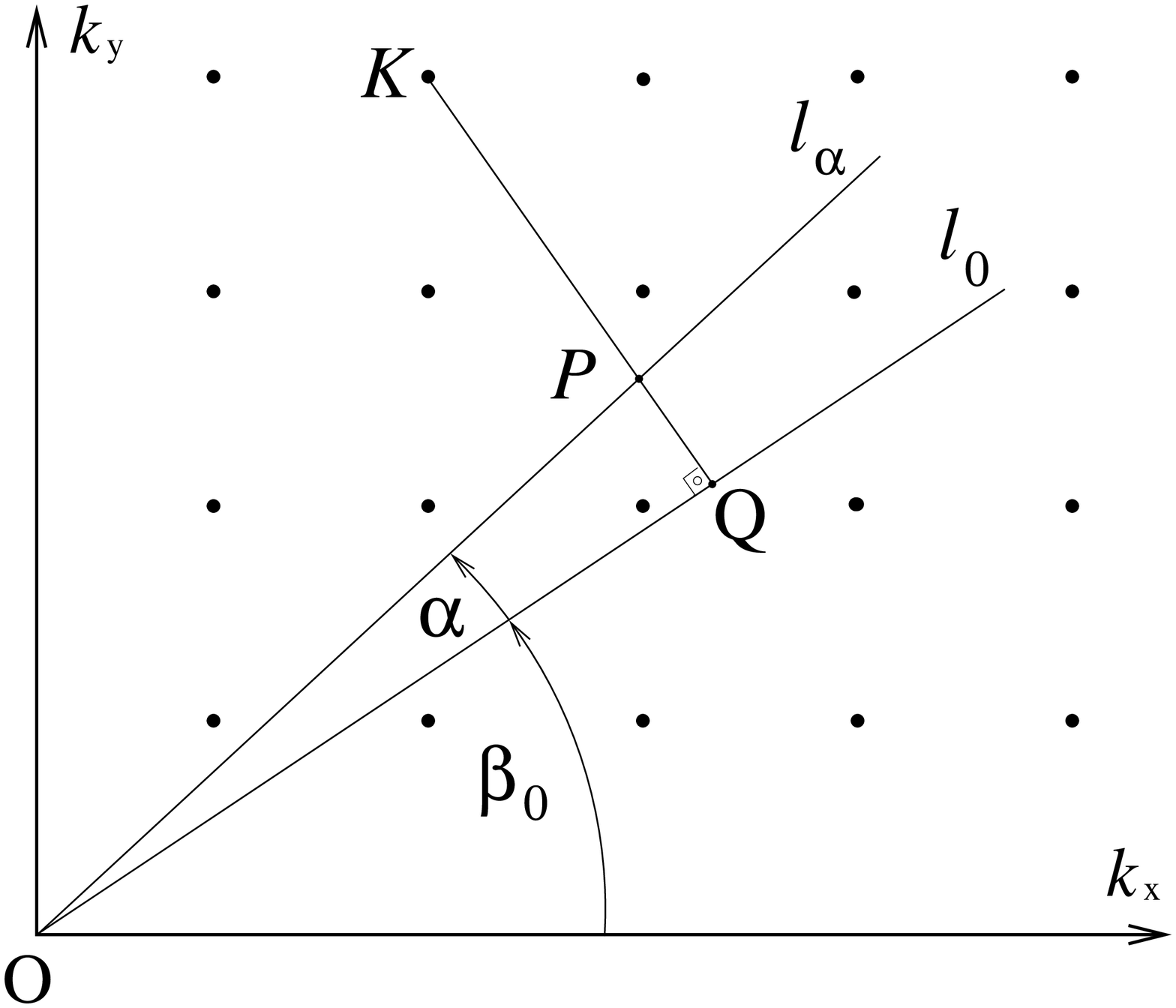}
        \end{center}
        \caption{Diffraction pattern of the modified Fibonacci structure is a section of the diffraction pattern
of $X$ through direction $l_{\alpha}$. $k^{phys}=\overline{OP}$ is the position of the peak, $\overline{KP}$
determines the intensity.}
        \label{fig:RecSpace}
\end{figure}

\section{Structure Factors and Average Unit Cells}
\label{sec:StructureFactor}
The concept of a reference lattice has previously been proposed in \cite{Wolny1998PhM}. Suppose we have a 1
dimensional Delone set $\left\lbrace r_{n} \right\rbrace$. Its points represent atoms, whose scattering
instensities are equal to unity. With some appropriate assumptions on the sequence $\{r_{n}\}$ we get the following 
expression for the structure factor:
\begin{align}
F(k) &= \lim_{N \to \infty} \frac{1}{N} \sum_{n=1}^{N} \exp (i k r_{n}) 
= \lim_{N \to \infty} \frac{1}{N} \sum_{n=1}^{N} \exp (i k u_{n}) \nonumber \\
&= \int_{- \lambda /2}^{\lambda /2} P (u) \exp (i k u) \textrm{d} u,
\label{eq:1pFac}
\end{align}
where $P(u)$ is the probability distribution of distances $u_{n}$ from the atoms to reference lattice positions
$m \lambda$, $\lambda=2 \pi /k$. That is,
\begin{align}
u_n=r_n - \left\| \frac{r_n}{\lambda} \right\| \lambda.
\end{align}

We call the series $u_n$ the \textit{displacements sequence} of $r_n$ (induced by the reference lattice with
period $\lambda$).  Any series $u'_n$ such that
\begin{equation}
u_n = u'_n - \left\| \frac{u'_n}{\lambda} \right\| \lambda,
\end{equation}
will be called an \textit{unreduced displacements sequence} (of $r_{n}$).

\newtheorem{theo}{Theorem}

\begin{theo}
\label{theo:sumOfDis}
Let $r_{n}=\alpha_n+\beta_n$ be a sum of two real series. If $d_{n}$ is a displacements sequence of $\alpha _{n}$
induced by a reference lattice, then
\begin{align}
u'_{n} &= d_{n} + \beta _{n} \label{eq:sumOfDis}
\end{align}
is an unreduced displacements sequence of $r_{n}$ induced by the same lattice.

\begin{proof}
Let $\lambda$ be the period of the reference lattice. We have to show that
\begin{align}
u_{n} = r_{n} - \left\| \frac{r_{n}}{\lambda} \right\| \lambda = u'_n - \left\| \frac{u'_n}{\lambda} \right\|
\lambda. 
\label{eq:toPr1}
\end{align}

Note that for any real number $x,y$, $\left \| x - \left\| y \: \right\| \right \| = \left\| x \right\| - \left\|
y \right\|$, since $\left\| y \right\|$ is an integer. We can write the right hand side of (\ref{eq:toPr1}) as
\begin{equation}
\beta_{n} + \alpha_{n} - \left\| \frac{\alpha _{n}}{\lambda} 
\right\| \lambda - \left\| \frac{\beta_{n} + \alpha_{n} - \left\| \alpha _{n} / 
\lambda \right\| \lambda}{\lambda} \right\| \lambda 
= r_{n} - \left\| \frac{r_{n}}{\lambda} \right\| \lambda,
\end{equation}
which is the left hand side.
\end{proof}
\end{theo}

The quantity $P(u)$ may be viewed as a probability distribution for an {\em average unit cell}.  The structure
factor for the scattering vector $k$ is just the first Fourier mode of this distribution.

Unfortunately, for each scattering vector we get, in principle, a different average unit cell and a different
distribution. However, the structure factor for $mk, \: m \in \mathbb{Z}$ can be computed from the reference
lattice for $k$; it is the $m$-th Fourier mode of the distribution $P(u)$. Thus, a single average unit cell is
sufficient to analyze structures whose scattering occurs at multiples of a fixed scattering vector $k_0$. This
situation includes, but is not limited to, the case where the original point pattern was periodic with period $2
\pi / k_0$.

For modulated structures (including quasicrystals), there are usually two periods, $a$ and $b$, which may be
incommensurate. Using two reference lattices, the first one having periodicity $a$ and the second having
periodicity $b$, the structure factor for the sum of two scattering vectors $k_{0} \equiv 2 \pi/ a$ and $q_{0}
\equiv 2 \pi/ b$ can be expressed by:
\begin{align}
F(k_{0} + q_{0}) &= \lim_{N \to \infty} \frac{1}{N} \sum_{n=1}^{N} \exp(i(k_{0} + q_{0})x_{n}) \nonumber \\
&= \lim_{N \to \infty} \frac{1}{N} \sum _{n=1} ^{N} \exp(i(k_{0} u_{n} + q_{0} v_{n})) \nonumber \\
&= \int _{-a/2} ^{a/2} \int _{-b/2} ^{b/2} P(u,v) \exp(i(k_{0} u +
q_{0} v)) \textrm{d}v \textrm{d}u, \label{eq:2pFac}
\end{align}
where $u$ and $v$ are the shortest distances of the atomic position from the appropriate points of two reference
lattices and $P(u,v)$ is the corresponding probability distribution, which thus describes a {\em two dimensional}
average unit cell. Likewise, the structure factor for a linear combination $nk_{0} + mq_{0}, \:n,\:m \in
\mathbb{Z}$ is given by the $(n,m)$ Fourier mode of $P(u,v)$. This means that the average unit cell, calculated
for the wave vectors of the main structure and its modulation, can be used to calculate the peak intensities of
any of the main reflections and its satellites of arbitrary order. Using (\ref{eq:2pFac}) and its generalization,
it is possible to calculate the intensities of all peaks observed in the diffraction patterns.

\section{1D Analysis of the Modified Fibonacci Chain}
\label{sec:MFCh1D}
Let $a=\frac{\tau + 2}{\tau + 1}$, the average spacing between atoms in a modified Fibonacci chain, and let $b =
\tau a$.  The positions of atoms in our chain are:
\begin{align}
  x_{m} &= \vec{p}_{m} \cdot (A,B) \nonumber \\
&= \left\| \frac{m}{\tau}\right\|A+mB - \left\|
          \frac{m}{\tau}\right\|B  \nonumber \\
&= (A-B)\left\|\frac{m}{\tau}\right\| + mB 
= (A-B)\Big[\frac{m}{\tau}+\frac{1}{2}\Big]+mB \nonumber \\
&= (A-B) \frac{m}{\tau} + mB + (A-B) \Big( \frac {1}{2} - 
\left\{ \frac{m}{\tau}+\frac{1}{2}\right\}\Big) \nonumber \\
&= ma+{u_{\scriptscriptstyle{0}}}M(ma), \label{eq:xmFin}
\end{align}
where $M(x)=\frac{1}{2}-\{\frac{x}{b}+\frac{1}{2}\}$ is a periodic function with period $b$ and
\begin{align}
u_{\scriptscriptstyle{0}}=\tau-1-\epsilon\tau. \label{eq:u0}
\end{align}
Here $[x] = \| x - \frac{1}{2} \|$ is the greatest integer function and $\{x\} = x - [x]$ is the fractional part
of $x$.

When $u_{\scriptscriptstyle{0}} \ne 0$, the modified Fibonacci chain is thus an incommensurately modulated
structure. A similar derivation for $\epsilon=0$ can be found in \cite{Senechal1995}.

We are going to construct a two dimensional average unit cell based on the two natural periodicities for $x_{m}$:
$a$ and $b$. For the scattering vector $k_0= 2 \pi / a$, it is obvious that the series $u'_{m} =
{u_{\scriptscriptstyle{0}}} M(ma)$ is an unreduced displacements sequence of $x_{m}$ induced by the reference
lattice with period $a$.

Next we consider the (one dimensional) average unit cell for the scattering vector $q_{0} = 2 \pi / b$. The
series
\begin{align}
\mu_{m} = ma-\left\| \frac{ma}{b} \right\| b = -b M(ma) = -u'_{m} \frac{b}{{u_{\scriptscriptstyle{0}}}}
\label{eq:nuseq}
\end{align}
is an (unreduced) displacement sequence of $ma$ induced by a reference lattice with period $b$.  Using Theorem
\ref{theo:sumOfDis} we immediately get that the series
\begin{align}
v'_{m} = \mu _{m} + u'_{m} = (u_{\scriptscriptstyle{0}}-b) M(ma) = v_{\scriptscriptstyle{0}} M(ma) = \xi u'_{m} \label{eq:v_prim}
\end{align}
is an unreduced displacements sequence of $x_{m}$ induced by a lattice with period $b$, where
\begin{equation}
{v_{\scriptscriptstyle{0}}}={u_{\scriptscriptstyle{0}}}-b.
\qquad 
\xi = {v_{\scriptscriptstyle{0}}}/{u_{\scriptscriptstyle{0}}} = 
\frac{-\tau^{2}(1+\epsilon)}{1-\epsilon\tau^{2}}. \label{eq:xi}
\end{equation}

By Kronecker's theorem (see \cite{Hardy1962}), the series $u'_{m}$ is uniformly distributed in the interval
$[-\left|{u_{\scriptscriptstyle{0}}}\right|/2, \left|{u_{\scriptscriptstyle{0}}}\right|/2]$. As pointed out by
Elser (\cite{Elser1986}, for a more precise discussion see also \cite{Senechal1995}) the uniformity of this
distribution is crucial for our deliberations. Likewise, the series $v'_{m}$ is uniformly distributed in the
interval $[-\left|{v_{\scriptscriptstyle{0}}}\right|/2,\left|{v_{\scriptscriptstyle{0}}}\right|/2]$.

The structure factor (see (\ref{eq:2pFac})) is
\begin{equation}
F(n_{1}k_{0} + n_{2}q_{0}) = \int _{-a/2} ^{a/2} \int _{-b/2} ^{b/2} P(u,v) \exp(i(n_{1}k_{0}u + n_{2}q_{0}v))
\textrm{d} v \textrm{d} u. 
\label{eq:factor}
\end{equation}
The unreduced displacements sequences $u'_{m}$ and $v'_{m}$ can be used to calculate $P(u,v)$. However, this
cannot be done directly because their terms may lie outside the average unit cell (i.e.:
$\left|u_{\scriptscriptstyle{0}}\right|>a$ or $\left|v_{\scriptscriptstyle{0}}\right|>b$). Such a situation is
shown in figure \ref{fig:AvCell1}. We have to reduce the series to the interior of the cell. The probability
function $P(u,v)$ is nonzero only along segments with slope $\xi$ (as a result of the strong correlation between
$u'_{m}$ and $v'_{m}$ given by (\ref{eq:v_prim})) and has constant value. This last fact follows from the
uniformity of the marginal distributions.

The formula (\ref{eq:factor}) is invariant under the changes
\begin{equation}
u \to u+\gamma_{1}a, \qquad
v \to v+\gamma_{2}b, \label{eq:inv} 
\end{equation}
where $\gamma_{1,2}$ are arbitrary integers. Likewise, the formula does not change if we use 
$P'(u,v)=C\delta(v-\xi u)$ instead of $P(u,v)$ and we change 
the area of integration from $[-a/2,a/2]$,$[-b/2,b/2]$ to 
$[-\left|{u_{\scriptscriptstyle{0}}}\right|/2,\left|{u_{\scriptscriptstyle{0}}}\right|/2]$, 
$[-\left|{v_{\scriptscriptstyle{0}}}\right|/2,\left|{v_{\scriptscriptstyle{0}}}\right|/2]$. That is, we are free
to integrate a part of distribution in the neighboring unit cells.

\begin{figure}[f]
        \begin{center}
        \includegraphics[width=0.75\textwidth,
        angle=0.0]{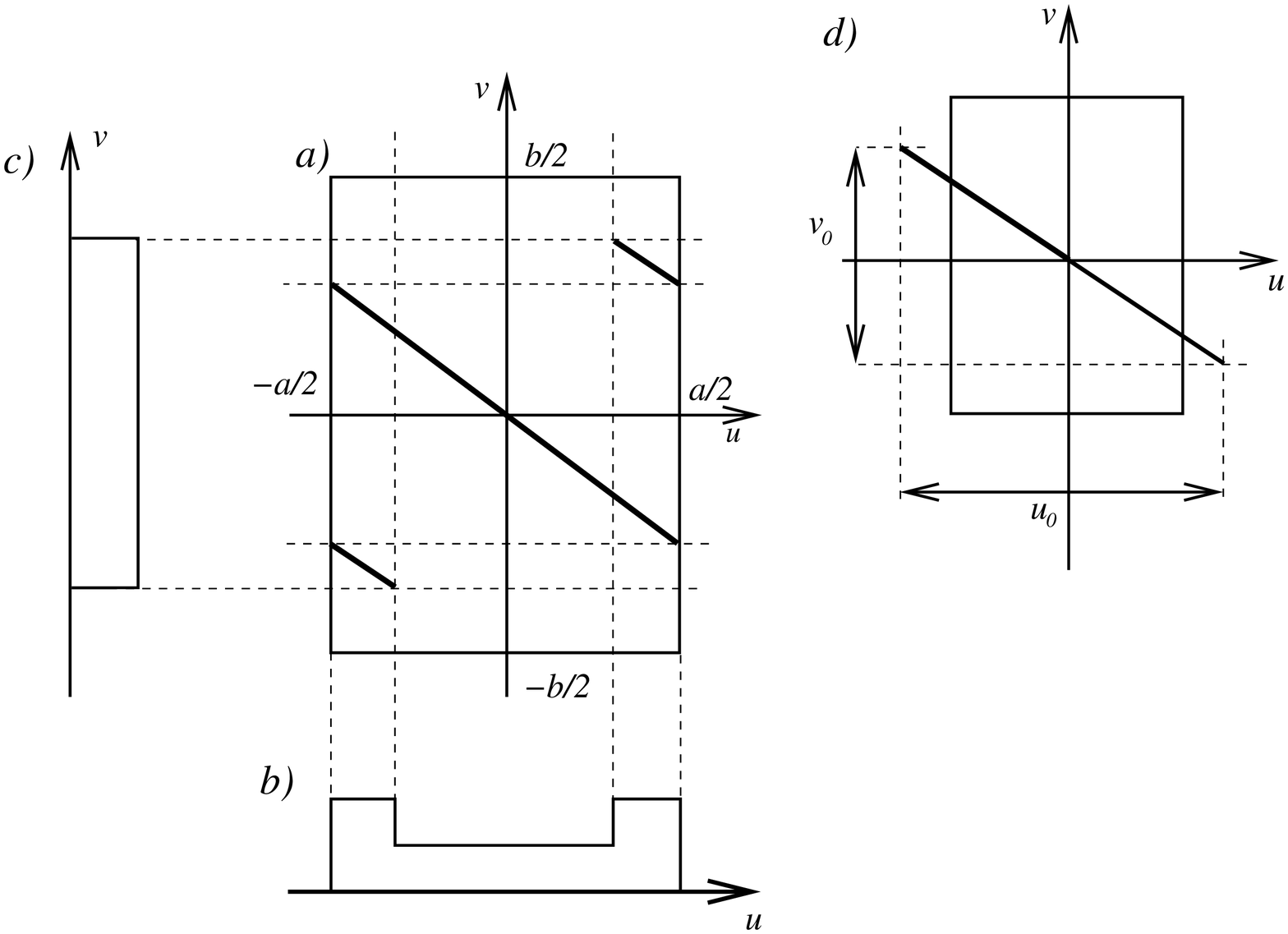}
        \end{center}
	\caption{(a) shows two parameter average unit cell.  The distribution $P(u,v)$ is non zero only along the
thick lines and has constant value. Projections of it onto directions $u$ and $v$ determine the probability
distributions for scattering vectors $k_{0}$ and $q_{0}$ ((b) and (c) respectively). (d) presents set of vectors
$\left\lbrace (u'_{m},v'_{m}) \mid m \in \mathbb{Z} \right\rbrace$. It may happen its elements lie outside the
average unit cell and have to be reduced to its interior (like for the presented example with $\epsilon = -0.7$).
Invariance under the substitution (\ref{eq:inv}) assures that integration of functions on (a) and (d) gives the
same results. Our parameter space has discrete translational symmetry like a periodic crystal.}
	\label{fig:AvCell1}
\end{figure}

For integers $n_1, n_2$ we compute the location $K_{n_1,n_2}$, the structure factor $F(K_{n_1,n_2})$ and the
normalized intensity $I$ of the corresponding peak:
\begin{equation}
 K_{n_{1}n_{2}} =(n_{1}k_{0}+n_{2}q_{0}) = \frac{2 \pi (\tau n_{1} + n_2)}
{\tau a} \qquad 
F(K_{n_1,n_2})  = \frac{\sin(w)}{w}, \qquad I = \left| F \right|^2, 
\label{eq:FinK}
\end{equation}
where
\begin{equation}
w  = (n_{1}k_{0}+n_{2}q_{0}\xi){u_{\scriptscriptstyle{0}}}/2 = 
(K_{n_1,n_2}-n_{2}q_{1}){u_{\scriptscriptstyle{0}}}/2, \qquad
  q_{1}=q_{0} (1 - \xi) = \frac{2 \pi \tau}{1 - \epsilon \tau^2}.
   \label{eq:Fin_w} 
\end{equation}

The integers $n_{1}$ and $n_{2}$ label the main reflection and its satellites, respectively. Equations
(\ref{eq:FinK}, \ref{eq:Fin_w})  can be used to calculate the positions and intensities of all peaks.

The correspondence with the previous 2 dimensional calculation is given by
\begin{equation}
n_{1}=m_{y}, \qquad
n_{2}=m_{x} - m_{y}. \label{eq:m_to_n}
\end{equation}
By equations (\ref{eq:10}) and (\ref{eq:FinK}), the peaks are located at
\begin{equation}
K_{n_{1},n_{2}} = K_{m_{y}, m_{x}-m_{y}} = 2 \pi \frac{m_y \tau + m_x - m_y}
{\tau} \frac{\tau + 1}{\tau + 2} = 2 \pi \frac{m_x \tau + m_y}{\tau + 2}
= k^{phys}.
\end{equation}
Likewise, substituting (\ref{eq:m_to_n}) into (\ref{eq:Fin_w}) and simplifying yields (\ref{eq:intens_2D}). It must be noted that the 2D approach presented here is not new and has been already used to describe the transformation between
quasiperiodic and periodic structures. Please refer to \cite{Kramer1987} and \cite{Torres1989}.

\section{Discussion of the Results}
\label{sec:Discussion}

It has been shown that the deformation rule in physical space changes only the amplitude of modulation (equation
(\ref{eq:xmFin})). Positions of peaks do not depend on the parameter $\epsilon$; only their intensities vary.
Using equations (\ref{eq:FinK}) and (\ref{eq:Fin_w}) we can easily build envelope functions, which go through the
satellite reflections of the same order (indexed by $n_{2}$). The shift of the envelope functions is $q_1$, as
given by (\ref{eq:Fin_w}).

The set of positions of Bragg peaks is always periodic, since the spectrum of a one-dimensional dynamical system
is an Abelian group.  By a \textit{commensurate} diffraction pattern we mean a pattern in which the amplitudes
are periodic as well. However, aside from the special case $A=B$, the Bragg peaks are described by two
incommensurate periods, and should not be confused with the diffraction of a periodic crystal.  For our
diffraction patterns, one period (of length $q_1$) is connected with envelope functions, while the second, with
period $k_0$, is associated with peaks ascribed to each envelope function. This behavior is characteristic of
modulated crystals and was discussed in \cite{Wolny1995}. Only the first periodicity could assure equality of
intensities.

It is convenient to describe our results in terms of the ratio $\kappa=A/B$. When $\kappa$ is rational, say equal
to $p/q$, then every atomic location $x_m$ is a multiple of $A/p=B/q$. Plane waves whose frequencies are
multiples of $2 \pi p/A$ have value 1 at each atomic position, and the entire diffraction pattern is periodic
with period $2 \pi p/A$.

Conversely, if the diffraction pattern is periodic, then the underlying periods $k_0$ and $q_1$ must be
commensurate.  A simple algebraic calculation then shows that $\kappa$ must be rational.  Thus, the pattern is
commensurate if and only if $\kappa$ is rational, which corresponds to projecting the set $X$ onto a rational
direction ($\kappa=\cot \beta$; see (\ref{eq:ABang})). 

Figure \ref{fig:ModFib0}a shows the diffraction pattern of an unmodified Fibonacci chain. The pattern is clearly
non-periodic, as $A/B$ equals to $\tau$. For $\epsilon=0$ our approach is identical with that presented in
\cite{Wolny1998ACr}. Figure \ref{fig:ModFib0}b shows the pattern for $\epsilon=-0.7$ (corresponding to the
average unit cell presented in figure \ref{fig:AvCell1}). As we can see, analytical calculations of envelope
functions are in full agreement with numerical calculations of the diffraction pattern.

\begin{figure}[f]
        \begin{center}
        \includegraphics[width=0.75\textwidth,
        angle=0.0]{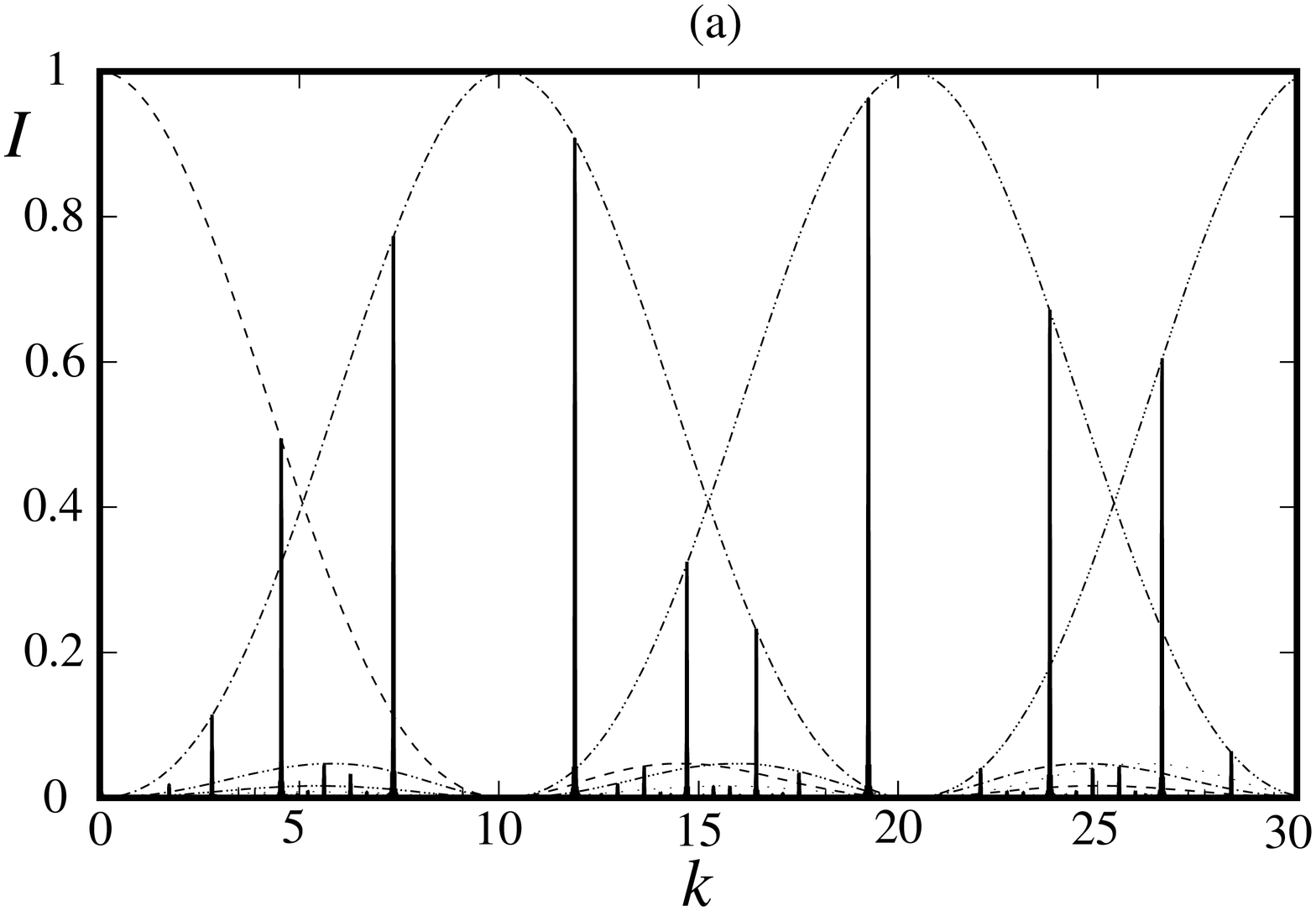}
        \includegraphics[width=0.75\textwidth,
        angle=0.0]{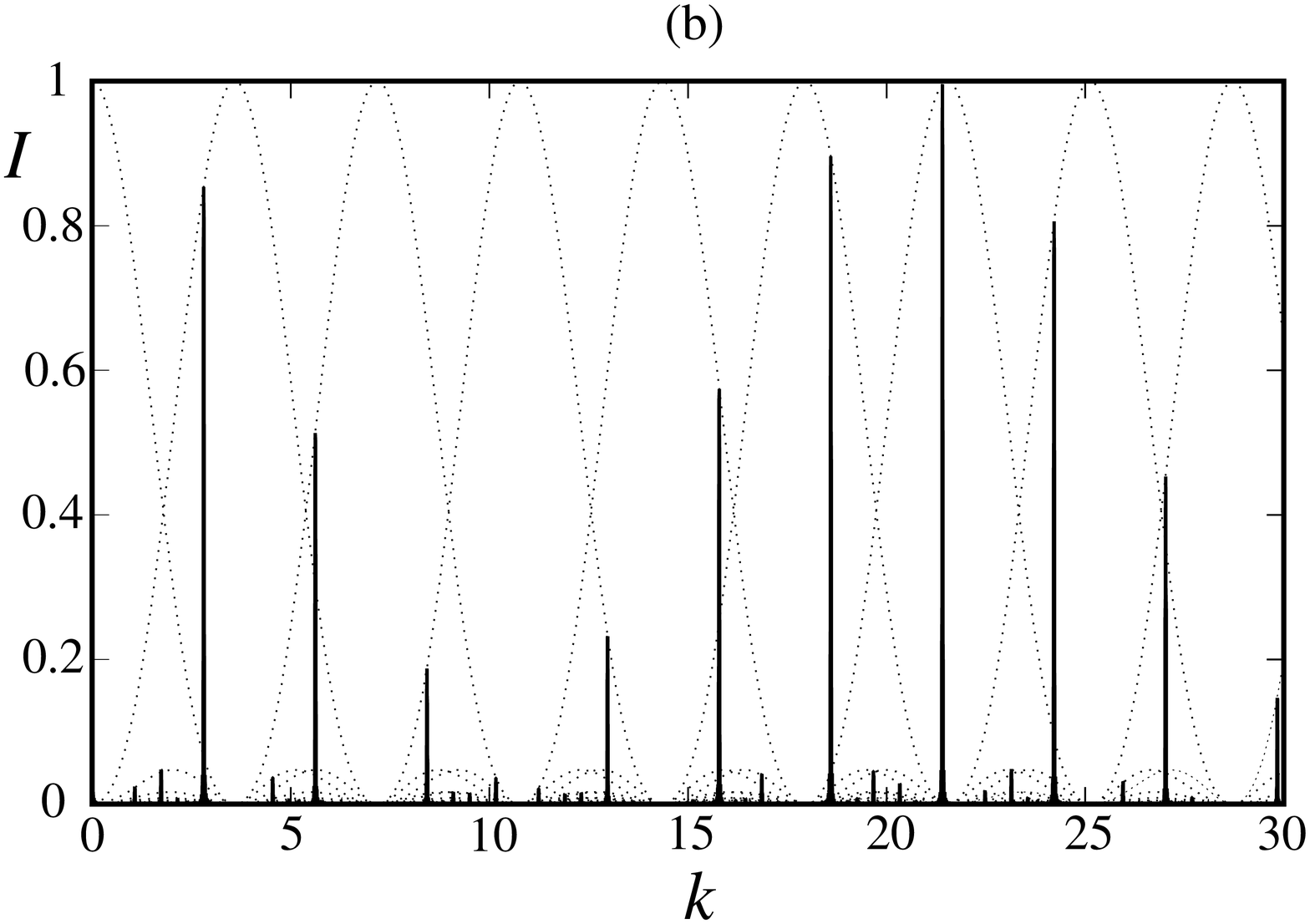}
        \end{center}
        \caption{Examples of diffraction patterns: (a) unmodified Fibonacci chain ($\kappa = \tau$ and $\epsilon
= 0$); (b) $\kappa \approx 6.836$ ($\epsilon = -0.7$). Broken lines present envelope functions. Given envelope
function goes through satellite peaks of the same order. All envelopes have the same shape; their shift is
$q_{1}$. As we can see analytical results are in full compatibility with numerical calculations.}
        \label{fig:ModFib0}
\end{figure} 

Figures \ref{fig:ModFib5} and \ref{fig:ModFib6} show diffractions patterns for $\kappa$ equal to $2/3$ and $3/2$
respectively. The regular series of peaks are clearly visible, but the diffraction patterns are still
quasicrystaline.  It is significant that for any value of $\epsilon$ except $1-1/\tau$ (discussed below) the
structure is not periodic in physical space, but may have periodic diffraction patterns.

\begin{figure}[f]
        \begin{center}
        \includegraphics[width=0.8\textwidth,
        angle=0.0]{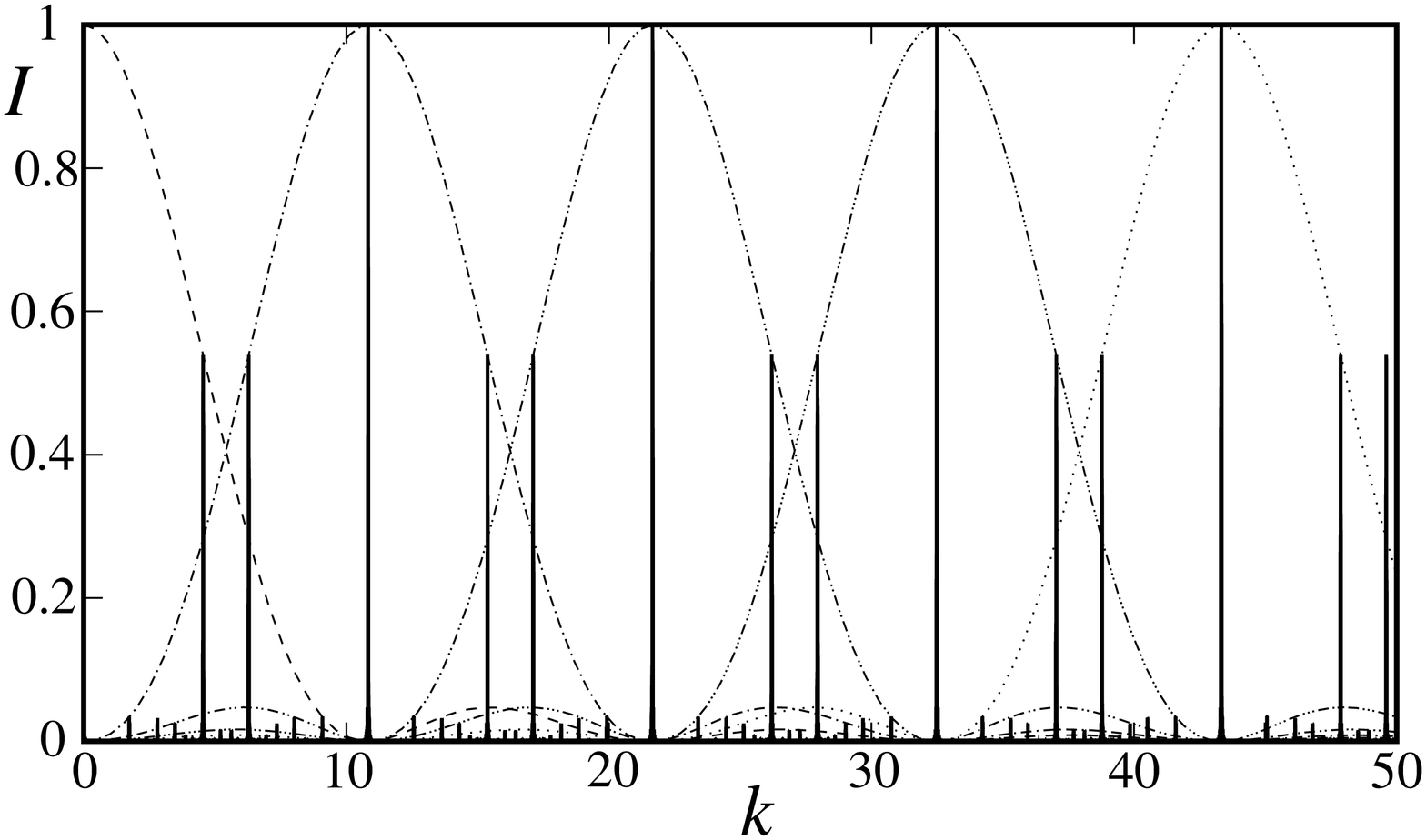}
        \end{center}
        \caption{Modified Fibonacci chain for $\kappa=2/3$ ($\epsilon \approx 0.741$, $q_{1} \approx 10.816$).}
	\label{fig:ModFib5}
\end{figure} 

\begin{figure}[f]
        \begin{center}
        \includegraphics[width=0.8\textwidth,
        angle=0.0]{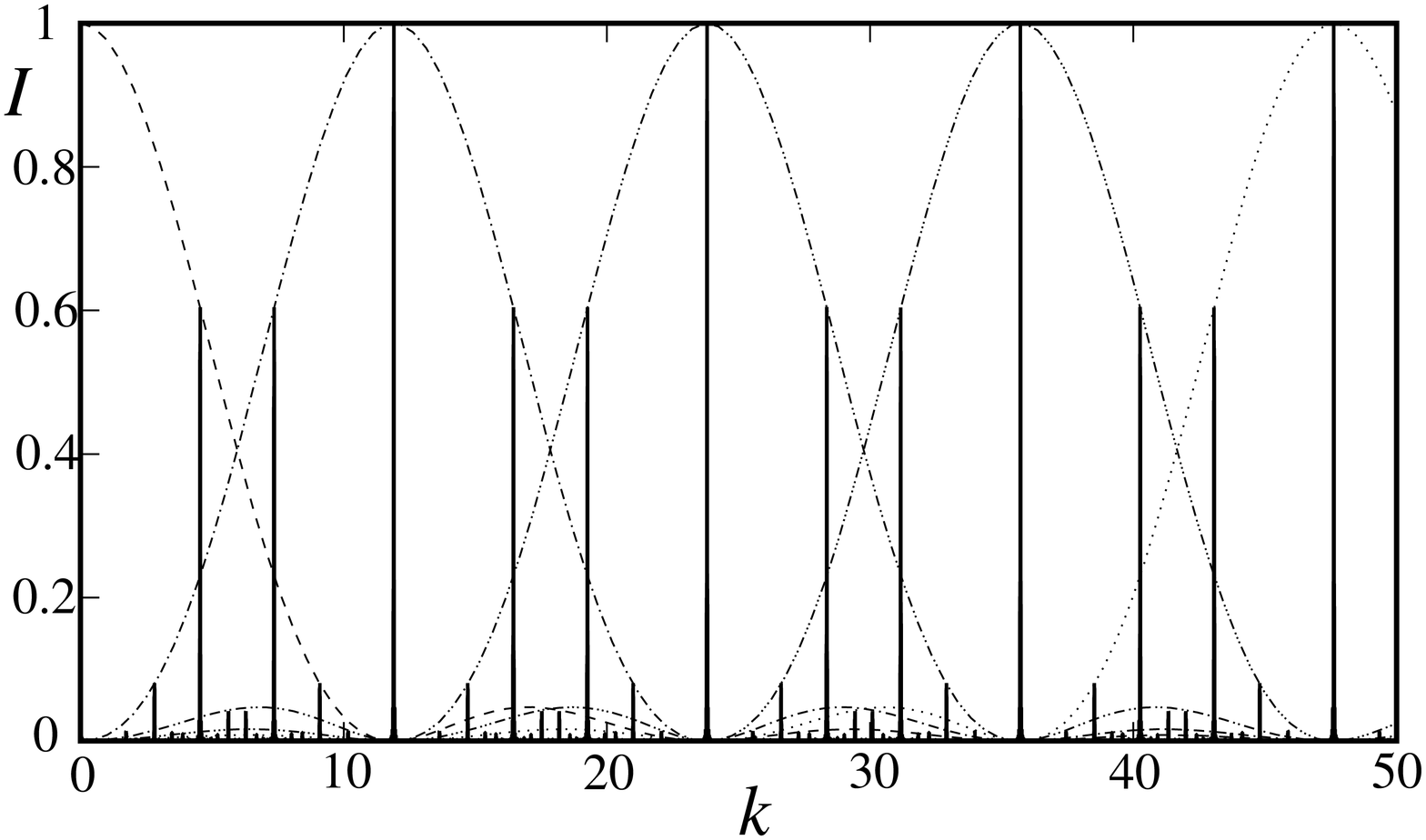}
        \end{center}
        \caption{Modified Fibonacci chain for $\kappa=3/2$ ($\epsilon \approx 0.056$, $q_{1} \approx 11.913$).}
        \label{fig:ModFib6}
\end{figure}

For $\epsilon=1 - 1 / \tau$ one gets fully periodic structure with $\kappa=1$, hence $A=B$.  Our deformed
Fibonacci chain is then simply a lattice, and its diffraction pattern is the reciprocal lattice, with period
$k_{0}$.  At this special value of $\kappa$, all the other Bragg peaks have intensity zero.

It must be also noted that the amplitude of each peak is a continuous function of $\kappa$. In fact, it is
infinitely differentiable. As $\kappa$ is varied, there is no phase transition between commensurate and
incommensurate diffraction patterns; the evolution is smooth. As such, with measurement apparatus of fixed
accuracy, it is impossible to determine whether a given pattern is precisely periodic.

Finally is may be noted that our analysis of the transition quasiperiodic-periodic (commensurate-incommensurate) is based on the explicit knowledge of the structure factor. It has been already shown (\cite{Kozakowski2003}) that
the method advertised (average unit cell) can give the factor for higher dimensional structures and therefore there are (in general) no obstacles to repeat similar analysis in the latter case. Such calculations have not been undertaken so far but it seems that they might be based on analytical expressions for the coordinates of quasiperiodic
lattices, derived from periodic or quasiperiodic grids (as given for example in \cite{Naumis2003}).

\listoffigures

\end{sloppy}
\end{document}